\documentstyle[12pt,epsfig,a4]{article} 
\newcommand{\vs}{\vspace{-0.25cm}}
\begin{document} 

\begin{center}
{\Large{\bf Chiral condensate in neutron matter}
}\footnote{Work supported in part by BMBF, GSI and by the DFG cluster of
excellence: Origin and Structure of the Universe.}

\medskip

N. Kaiser and W. Weise\\

\smallskip

{\small Physik Department, Technische Universit\"{a}t M\"{u}nchen,
    D-85747 Garching, Germany}
\end{center}

\medskip

\begin{abstract}
A recent chiral perturbation theory calculation of the in-medium quark 
condensate $\langle \bar q q\rangle$ is extended to the isospin-asymmetric case
of pure neutron matter. In contrast to the behavior in isospin-symmetric 
nuclear matter we find only small deviations from the linear density 
approximation. This feature originates primarily from the reduced weight 
factors (e.g. $1/6$ for the dominant contributions) of the $2\pi$-exchange 
mechanisms in pure neutron matter. Our result suggests therefore that the 
tendencies for chiral symmetry restoration are actually favored in systems
with large neutron excess (e.g. neutron stars). We also analyze the behavior 
of the density-dependent quark condensate $\langle  \bar q q\rangle(\rho_n)$
in the chiral limit $m_\pi\to 0$.  
\end{abstract}
\medskip 

\section{Introduction and framework} 
The quark condensate $|\langle 0|\bar q q|0\rangle|$ is an order parameter of 
spontaneous chiral symmetry breaking in QCD. With increasing temperature the 
quark condensate decreases (or ``melts''). For low temperatures $T$ this 
effect can be systematically calculated in chiral perturbation theory. The
estimate $T_c \simeq 190\,$MeV  \cite{gerber,garcia} for the critical 
temperature, where chiral symmetry will eventually be restored, has been
found. This extrapolated value is remarkably consistent with $T_c = (192\pm
8)\,$MeV \cite{cheng} obtained in recent lattice QCD simulations (modulo still 
persisting disputes between different lattice groups \cite{aoki}). 

The chiral condensate  $|\langle\bar q q\rangle|$ drops also with increasing 
baryon density. Presently, it is not feasible to study this phenomenon
rigorously in lattice QCD due to the problems arising from the 
complex-valued fermion determinant at non-zero quark chemical potential. As 
an alternative, the density dependence of $\langle\bar q q\rangle(\rho)$ can 
be extracted by exploiting the Feynman-Hellmann theorem applied to the chiral 
symmetry breaking quark mass term $m_q\, \bar qq$ in the QCD Hamiltonian. The 
leading linear term in the nucleon density $\rho$ is then readily derived by 
differentiating the energy density of a nucleonic Fermi gas, $\rho M_N + {\cal 
O}(\rho^{5/3})$, with respect to the light quark mass $m_q$. This introduces the 
nucleon sigma-term $\sigma_N = \langle N|m_q\,\bar q q|N\rangle = m_q\,
\partial M_N /\partial m_q =(45\pm 8) \,$MeV \cite{sigma} as  the driving term 
for the density evolution of the chiral condensate. Following this leading
linear density approximation one would estimate that chiral symmetry gets 
restored at $(2.5-3)\rho_0$, with $\rho_0 = 0.16\,$fm$^{-3}$ the nuclear
matter  saturation density. 

Corrections beyond the linear density approximation arise from the
nucleon-nucleon correlations which transform the nucleonic Fermi gas into a 
nuclear Fermi liquid. Because of the Goldstone boson nature of the pion, with 
its characteristic mass relation $m_\pi^2 \sim m_q$, the pion-exchange dynamics
in nuclear matter plays a particularly important role for the in-medium quark 
condensate. In a recent work \cite{homont} we have used in-medium chiral
perturbation theory to calculate systematically the corrections to the 
linear density approximation. This calculation has treated in detail the 
long- and medium-range correlations arising from $1\pi$-exchange (with 
$m_\pi$-dependent vertex corrections), iterated $1\pi$-exchange, and
irreducible $2\pi$-exchange including also virtual $\Delta(1232)$-isobar
excitations, with Pauli-blocking corrections up to three-loop order. It was
furthermore necessary  to estimate the quark mass dependence of a NN-contact
term which encodes unresolved short-distance dynamics. Employing a recent
computation of the NN-potential in lattice QCD \cite{hatsuda} (at three 
different pion masses) we have found that the contact term has a negligible 
influence on  the in-medium quark condensate (in agreement with 
ref.\cite{fuchs} which follows a somewhat different approach). As a result, we 
have obtained in  ref.\cite{homont} a strong and non-linear dependence of the 
``dropping'' condensate on the actual value of the pion mass $m_\pi$. In the
chiral limit $m_\pi=0$, chiral symmetry would seem to be restored already at 
about $1.5\rho_0$. By contrast, for the physical pion mass $m_\pi=135\,$MeV, the
in-medium condensate stabilizes at about $60\%$ of its vacuum value around that
same density  (see Fig.\,2 in the present paper). 

Having found such pronounced deviations from the linear density approximation, 
with a hindered tendency towards chiral symmetry restoration in 
isospin-symmetric nuclear matter, it is a logical next step to explore the 
effects of the additional isospin degree of  freedom present in nuclear 
many-body systems. Such a novel study is the purpose of the present short 
paper where we consider the extreme isospin-asymmetric case of pure neutron 
matter. Interestingly, we find that the same chiral pion-exchange dynamics 
which stabilizes the quark condensate in isospin-symmetric nuclear matter does 
not alter (in a significant way) its linear decrease in pure neutron matter
(for densities $\rho_n <0.35\,$fm$^3$). At the same time the behavior in the
chiral limit $m_\pi \to 0$ changes prominently once a nonvanishing
isospin-asymmetry is present in the nuclear medium.         
          
As in ref.\cite{homont}, our starting point is the Feynman-Hellmann theorem. 
It relates the in-medium quark condensate $\langle \bar q q\rangle(\rho_n)$ to 
the quark mass derivative of the energy density of pure neutron matter. Using
the Gell-Mann-Oakes-Renner relation $m_\pi^2 f_\pi^2 =-m_q \langle 0|\bar q q|0 
\rangle$ one obtains for the ratio of the in-medium to the vacuum quark
condensate: 
\begin{equation} { \langle \bar q q\rangle(\rho_n)\over \langle 0|\bar q q|0 
\rangle} = 1 -{\rho_n \over f_\pi^2} \bigg\{ {\sigma_N \over m_\pi^2} \bigg(1 
-{3k_n^2 \over 10 M_N^2}+{9k_n^4\over 56 M_N^4}  \bigg) + D_n(k_n) \bigg\}\,,
\end{equation}
with $k_n$ the neutron Fermi momentum and $\rho_n = k_n^3/3\pi^2$ the neutron
density. The term proportional to $\sigma_N = \langle N|m_q\,\bar q q|N\rangle 
= m_q\,\partial M_N /\partial m_q$ comes from the non-interacting Fermi gas
including the (relativistically improved) kinetic energy. We mention here that 
$f_\pi$ denotes the pion decay constant in the chiral limit and $m_\pi^2$ stands 
for the leading linear term in the quark mass expansion of the squared pion
mass. Interaction contributions beyond the linear density approximation are 
collected in the function:    
\begin{equation}D_n(k_n) =  {\partial \bar E_n(k_n) \over \partial  m_\pi^2}  \,, 
\end{equation}
defined as the derivative of the interaction energy per particle $\bar
E_n(k_n)$ with respect to $m_\pi^2$. Note that small (explicit) isospin breaking 
effects \cite{mow} (in interactions etc.) are not included in our calculation. 
The asymmetry in isospin is entirely given by the filled Fermi sea of neutrons 
(and the empty one for protons).    

\section{Selected interaction contributions}
In this section we present analytical results for the contributions to the 
function $D_n(k_n)$ as given by various classes of $1\pi$- and $2\pi$-exchange 
diagrams, calculated up to three-loop order in the energy density $\rho_n \bar 
E_n(k_n)$. At the level of individual diagrams pure neutron matter and
isospin-symmetric nuclear matter differ only with respect to the (overall)
isospin factor. In Table\,I several of these relative isospin factors are 
listed. Therein, we refer to the corresponding equation for $D(k_f)$ in 
ref.\cite{homont} and prescribe to substitute the Fermi momentum: $k_f\to k_n$.
For completeness and clearness the remaining $2\pi$-exchange contributions to 
$D_n(k_n)$ are better written out explicitely. We are following the
enumeration scheme introduced in Sec.\,II of ref.\cite{homont}.
\vskip -0.4cm
\begin{table}[hbt]
\begin{center}
\begin{tabular}{|c|ccccccccc|}
\hline 
Eq. in ref.\cite{homont}& (4) & (5) & (6) & (7) & (8) & (9) & (12) & (22) & 
(23)  \\ \hline
isospin factor & $1/3$ & $1/3$  & $1/6$  & $-1/3$ & $1/6$  & $-1/3$ & $1/6$ &  
$1/6$  & $1/3$   \\
\hline
\end{tabular}
\end{center}
\vskip -0.1cm
{\it Table\,I: Relative isospin factors for several $1\pi$- and 
$2\pi$-exchange contributions to  $D_n(k_n)$.}
\end{table}
\vskip -0.2cm

The irreducible $2\pi$-exchange with only nucleons in the intermediate state
leads to the following contribution \cite{nucmat}:
\begin{eqnarray} 
D_n(k_n)^{(2\pi)}&=& {m_\pi^3\over (4\pi f_\pi)^4} \Bigg\{ \bigg[ {1\over 8u^3} 
(83g_A^4+6g_A^2-1)+{3\over 4u}(47g_A^4+2g_A^2-1)\bigg] \ln^2(u+\sqrt{1+u^2})
\nonumber \\ && +\bigg[{1\over 4u^2}(1-6g_A^2-83g_A^4)-{4\over 3}-2g_A^2+{86
\over 3}g_A^4-{u^2\over 3}(g_A^4+6g_A^2+1)\bigg]\sqrt{1+u^2}\nonumber\\ &&\times 
 \ln(u+\sqrt{1+u^2}) +{1\over 8u}(83g_A^4+6g_A^2-1)+{u\over 24}(47+30g_A^2-1285
g_A^4)\nonumber \\&&  +{u^3\over 12} (9+46g_A^2-55g_A^4)+{u^3\over 3}(1+6g_A^2-
15g_A^4)\ln{m_\pi\over \lambda} \Bigg\} \,,  \end{eqnarray}
with $u = k_n/m_\pi$ and $\lambda$ the regularization scale. The three-body
Fock term related to $2\pi$-exchange with virtual $\Delta$-excitation reads: 
\begin{equation}
D_n(k_n)^{(\Delta,F3)} = {g_A^4 m^4_\pi \over 4\Delta (4\pi f_\pi)^4 u^3}
\int_0^u \!\!dx\bigg\{G_S \bigg( x {\partial  G_S \over \partial x} +u 
{\partial G_S \over \partial u}-4G_S  \bigg) +2G_T\bigg( x {\partial  G_T \over 
\partial x}+u {\partial G_T\over \partial
u}-4G_T\bigg)\bigg\}\,,\end{equation}
with the auxiliary functions $G_{S,T}(x,u)$ written in Eqs.(7,8) of
ref.\cite{deltamat}.  The dominant two-body terms scaling reciprocally with the
delta-nucleon mass splitting $\Delta = 293\,$MeV take the form:
\begin{eqnarray}
D_n(k_n)^{(\Delta2)} &=& {\pi g_A^4m^4_\pi \over 70\Delta(2\pi f_\pi)^4}\bigg\{
( 140+42u^2 +15u^4) \arctan u \nonumber \\ && -{89+315u^2\over 4u^3}\ln(1+u^2)  
 +{89 \over 4u } -{579 u \over 8} -{1115 u^3 \over 12} 
\bigg\}\,. \end{eqnarray}
For the remaining two-body terms with a more complicated $\Delta$-dependence
we employ the spectral-function representation \cite{deltamat} and 
differentiate directly the imaginary parts of the $\pi N\Delta$-loop functions 
with respect to $m_\pi^2$. This gives: 
\begin{eqnarray}
D_n(k_n)^{(\Delta2')}&=& {g_A^2\over (4\pi f_\pi)^4} \int_{2m_\pi}^\infty\!\!
d\mu \bigg[ 3\mu k_n -{4k_n^3 \over 3\mu} -{\mu^3 \over 2k_n} - 4 \mu^2 
\arctan{2k_n\over \mu} +{\mu^3 \over 8k_n^3}(12k_n^2+\mu^2)  \nonumber \\ &&
\times \ln\bigg( 1+{4k_n^2\over\mu^2}\bigg)\bigg] \Bigg\{\bigg[ {2 \Delta 
\over \mu} +{g_A^2 \over 8  \mu \Delta} (40 \Delta^2 +72m_\pi^2 -17 \mu^2) 
\bigg] \arctan{\sqrt{ \mu^2-4m_\pi^2}\over 2\Delta}\nonumber \\ &&-{2g_A^2 \mu 
m_\pi^2 \over \Delta^2 \sqrt{\mu^2- 4m_\pi^2}}+\sqrt{\mu^2-4m_\pi^2} \bigg[ 
{7g_A^2 \mu(m_\pi^2 - \Delta^2)\over(\mu^2+ 4\Delta^2-4m_\pi^2)^2} -{2+5g_A^2 
\over 2 \mu} \nonumber \\ &&+ {4g_A^2 \mu (m_\pi^2 -\Delta^2) -\mu \Delta^2 
\over 2\Delta^2(\mu^2+ 4\Delta^2-4 m_\pi^2)} \bigg]  \Bigg\}  \,,  
\end{eqnarray}
after subtracting a term linear in the neutron density $\rho_n = k_n^3/3\pi^2$. 
The associated subtraction constant includes also pion-loop contributions with 
a nonanalytical dependence on the quark mass $m_q$. We reinstore these 
distinguished pieces by the term: 
\begin{equation} D_n(k_n)^{(dt)} = {g_A^2 k_n^3\over (4\pi f_\pi)^4} \Bigg\{ 
(5g_A^2-2) \ln {m_\pi \over 2\Delta} +{5g_A^2(2\Delta^2-9m_\pi^2)-4\Delta^2
\over 2\Delta \sqrt{\Delta^2-m_\pi^2} }\ln {\Delta+  \sqrt{\Delta^2-m_\pi^2}
\over m_\pi} \Bigg\} \,. \end{equation}
Finally, there is the $2\pi$-exchange two-body term generated by the $\pi\pi
NN $-contact vertex proportional to the low-energy constant $c_1 = -\sigma_N/
4m_\pi^2+{\cal O}(m_\pi)$ (measuring explicit chiral symmetry breaking in the 
$\pi N$-interaction). Its contribution to the function $D_n(k_n)$ reads: 
\begin{equation}
D_n(k_n)^{(c_1,2)} = {3g_A^2 c_1m^4_\pi \over 280\pi^3 f_\pi^4}\bigg\{
(14u^2 +3u^4) \arctan u +{27+49u^2\over 4u^3}\ln(1+u^2)   -{27 \over 4u } 
-{71 u \over 8} -{93u^3 \over 4} \bigg\}\,. \end{equation}
In comparison to Eq.(20) in ref.\cite{homont} only the numerical coefficient
of the last $u^3$-term has changed. This comes from the different weighting of 
the Hartree and Fock contributions in neutron matter as compared to
isospin-symmetric nuclear matter. 

The sum of all terms, Eqs.(3-8) together with those obtained via the relative 
isospin factors in Table\,I, comprise the $m_\pi^2$-derivative of the complete 
set of in-medium $1\pi$- and $2\pi$-exchange processes up to three-loop order 
in the energy density (with inclusion of explicit $\Delta(1232)$ degrees of 
freedom).   

\section{Results and discussion}
We are using consistently the same parameters in the chiral limit as in our
previous work \cite{homont}, namely: $f_\pi = 86.5\,$MeV, $g_A=1.224$, $c_1
=-0.93\,$GeV$^{-1}$  and $M_N = \lambda= 882\,$MeV. For the quark mass 
dependence of the short-distance dynamics (not controlled by the underlying 
chiral effective field theory) we  adopt the result derived in 
ref.\cite{homont} via the short range part ($r\leq 0.6\,$fm) of the 
NN-potential from lattice QCD \cite{hatsuda}. Its effect on the in-medium  
condensate  $\langle \bar q q \rangle(\rho_n)$ is then again negligibly small. 
\begin{figure}
\begin{center}
\includegraphics[scale=0.5,clip]{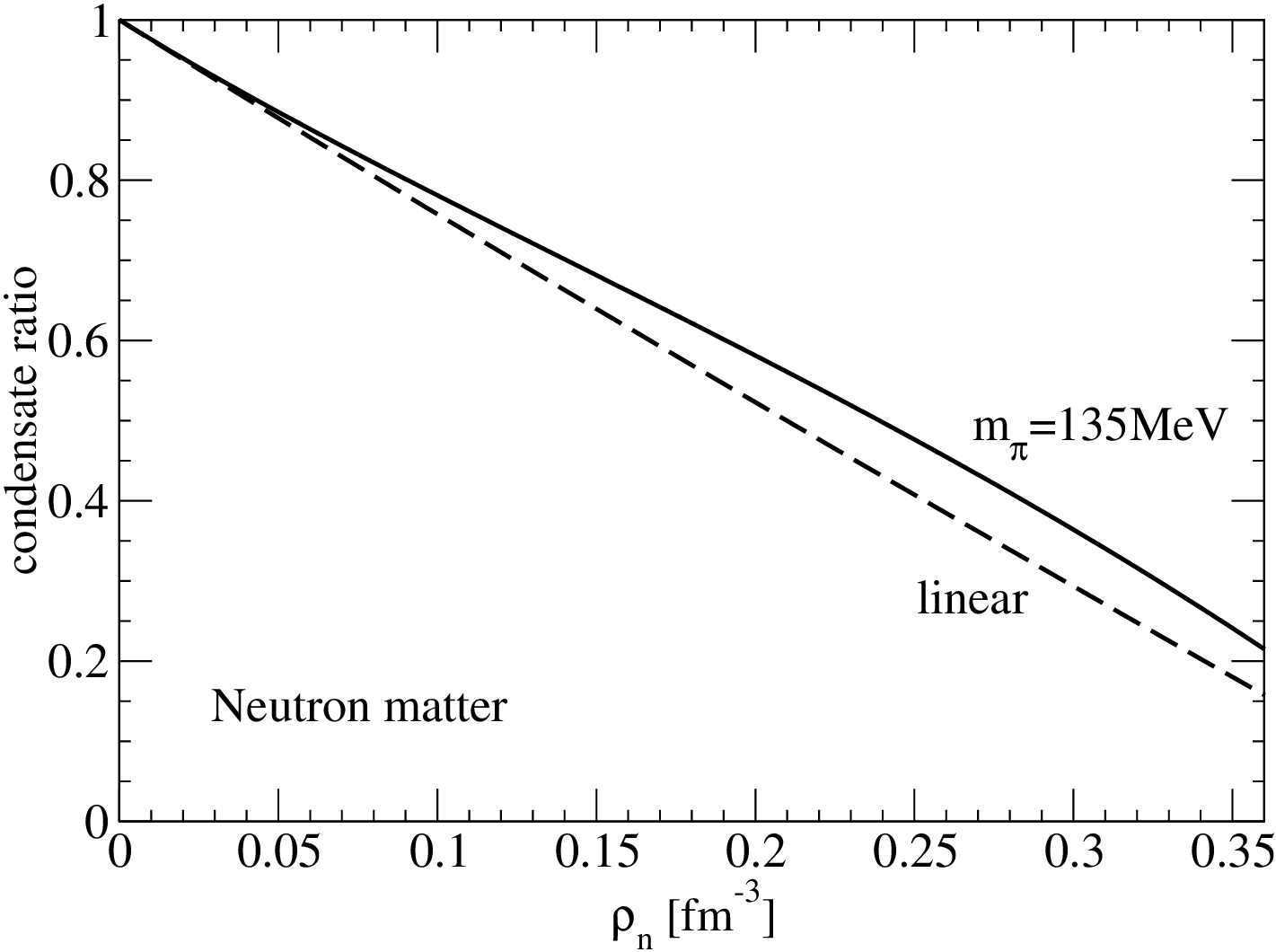}
\end{center}
\vspace{-0.4cm}
{\it Fig.\,1: Ratio of the in-medium chiral condensate in pure neutron matter
to its vacuum value at the physical pion mass, $m_\pi = 135\,$MeV. The dashed 
line corresponds to the linear density approximation using the empirical 
central value $\sigma_N = 45\,$MeV \cite{sigma}.} 
\end{figure}

Collecting all the pieces entering into Eq.(1), the condensate ratio 
$\langle \bar q q \rangle(\rho_n)/\langle 0|\bar q q|0\rangle$ in pure neutron 
matter comes out as shown by the full line in Fig.\,1. The dashed line therein
corresponds to the linear density approximation using the empirical central 
value of the nucleon sigma-term, $\sigma_N =45\,$MeV \cite{sigma}. In 
contrast to the behavior in isospin-symmetric nuclear matter (redisplayed in 
Fig.\,2 for comparison) the chiral pion-exchange dynamics in pure neutron 
matter generates only small deviations from the linear dropping of the 
in-medium condensate with density $\rho_n$. This feature is straightforwardly 
explained by the reduced isospin weight factors of the $2\pi$-exchange 
mechanisms ($1/6$ for the dominant contributions). 
Fig.\,3 shows separately the effects of the five classes of interaction 
contributions for neutron densities $\rho_n \leq 0.35\,$fm$^{-3}$. They are 
consecutively added up in the sequence: linear density approximation $\to
1\pi$-exchange $\to$ iterated $1\pi$-exchange $\to$ irreducible $2\pi$-exchange
$\to$  $2\pi$-exchange with virtual $\Delta(1232)$-excitation $\to$ chiral
symmetry breaking $c_1$-term. One observes that the effects from 
$2\pi$-exchange  cancel here almost completely such that the total result 
(full line) lies close to the $1\pi$-exchange approximation (dashed-dotted 
line). This behavior is markedly different from the situation in 
isospin-symmetric nuclear matter (see Fig.\,6 in ref.\cite{homont}). Note also 
that all the results discussed  so far refer to the physical value of the pion 
mass, $m_\pi =135\,$MeV. 

\begin{figure}
\begin{center}
\includegraphics[scale=0.5,clip]{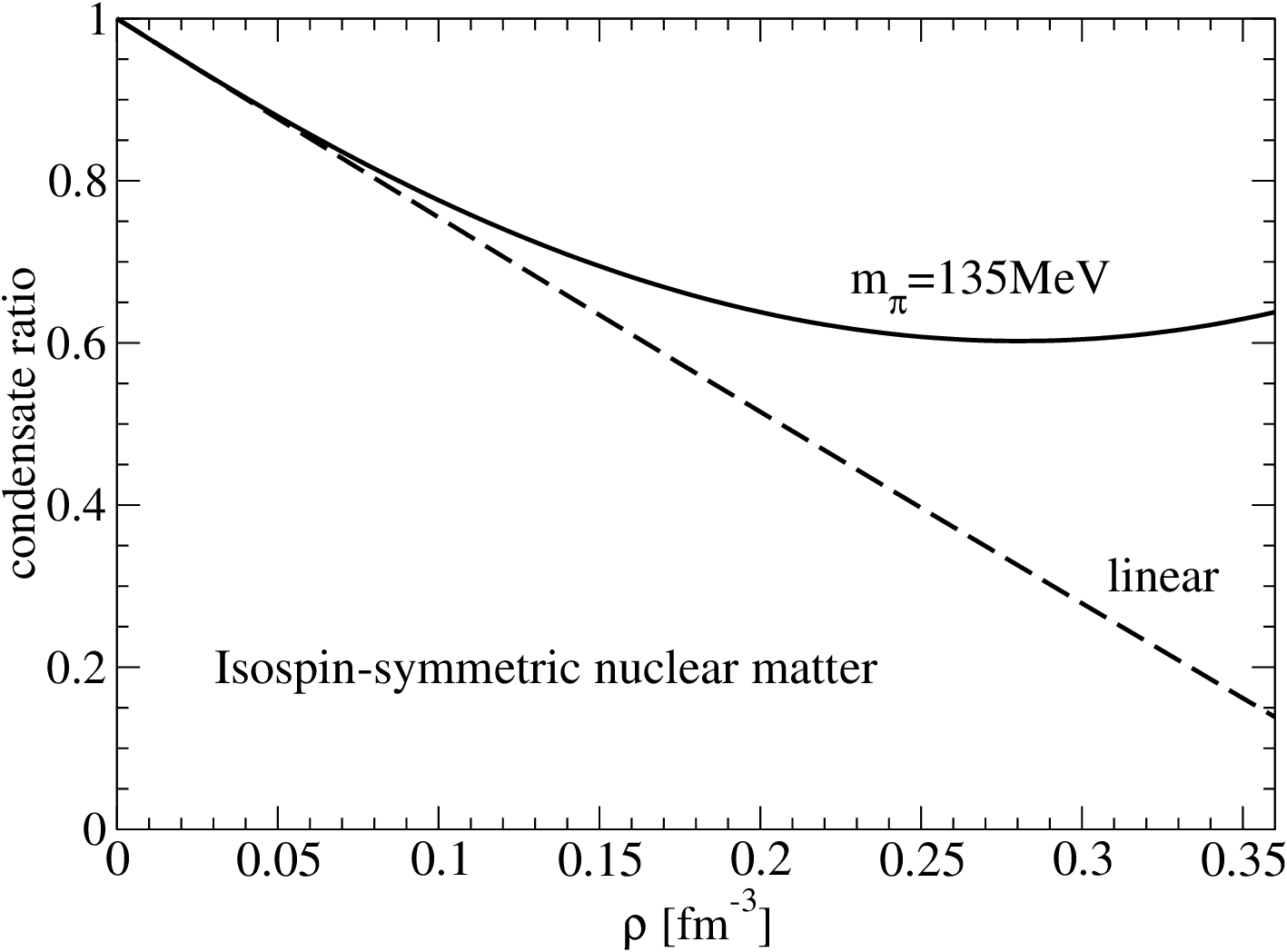}
\end{center}
\vspace{-0.4cm}
{\it Fig.\,2: Ratio of the in-medium chiral condensate in isospin-symmetric
nuclear matter to its vacuum value. The dashed line shows the linear density 
approximation.}
\end{figure}
\begin{figure}
\begin{center}
\includegraphics[scale=0.5,clip]{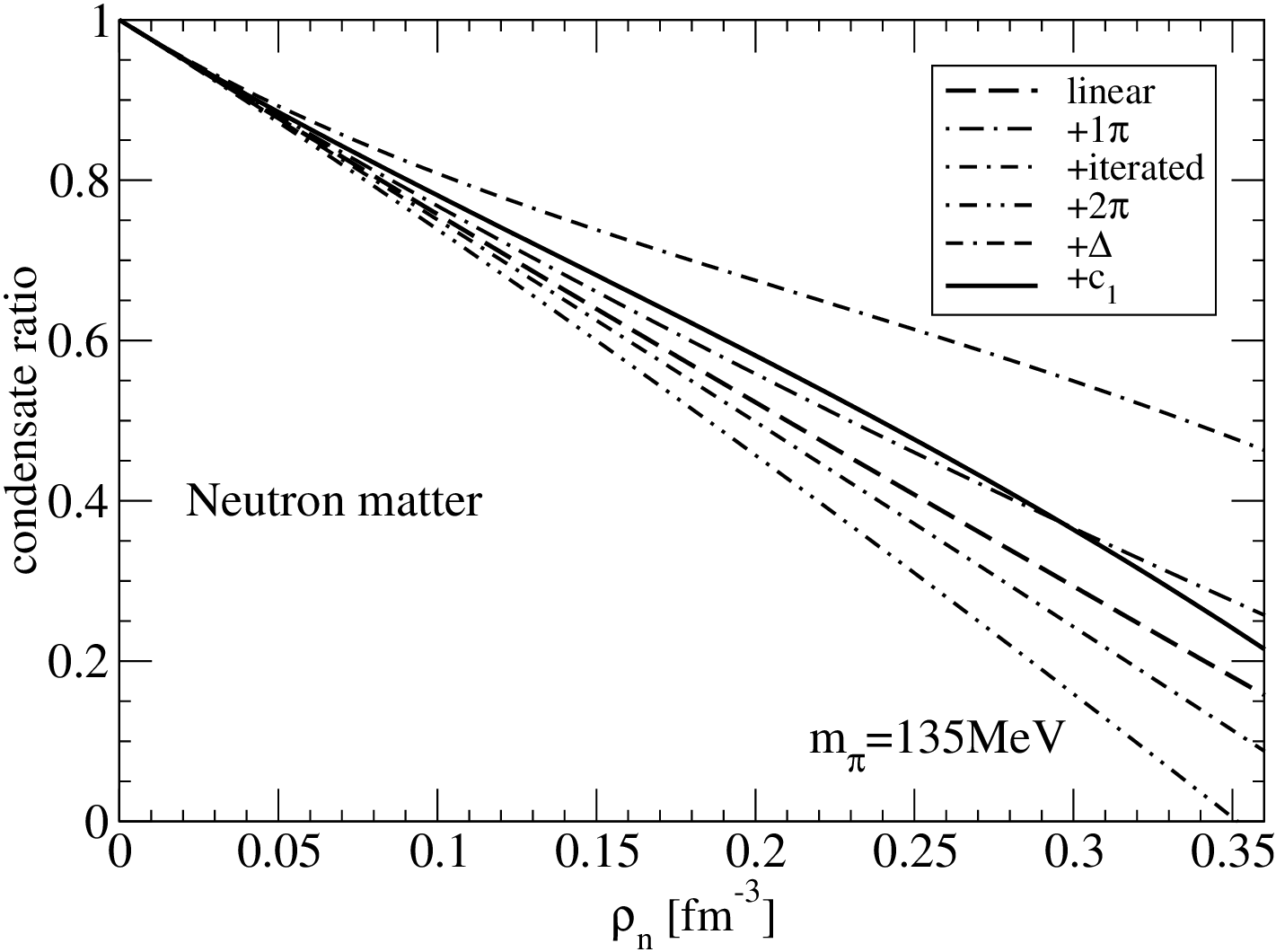}
\end{center}
\vspace{-0.4cm}
{\it Fig.\,3: Ratio between the in-medium chiral condensate in pure neutron 
matter and its vacuum value. The five classes of interaction contributions 
are  consecutively added in the sequence: linear $\to 1\pi \to$ iterated $\to 
2\pi \to \Delta \to c_1$.}
\end{figure}

For the sake of completeness, we show in Fig.\,4 the neutron matter equation 
of state resulting from our choice of parameters. Besides the $1\pi$- and
$2\pi$-exchange contributions described in refs.\cite{nucmat,deltamat} and
those proportional to $c_1$, it includes an adjusted short-distance term of
the form: $\bar E_n(k_n)^{(adj)} = -1.04\,$GeV$^{-2}\,k_n^3-18.4\,$GeV$^{-4}\,k_n^5$. 
For not too high neutron densities, $\rho_n \leq 0.2 \,$fm$^{-3}$, our 
perturbative calculation reproduces fairly well the result of the Urbana group 
\cite{akmal} based on a sophisticated many-body calculation. At higher neutron
densities we get (as in ref.\cite{deltamat}) a stiffer neutron matter equation
of state. This has to do with a repulsive $k_n^6$-term generated by the
$2\pi$-exchange three-neutron interaction. Note however that this $k_n^6$-term 
drops out  when taking the derivative with respect to $m_\pi^2$.
\begin{figure}
\begin{center}
\includegraphics[scale=0.5,clip]{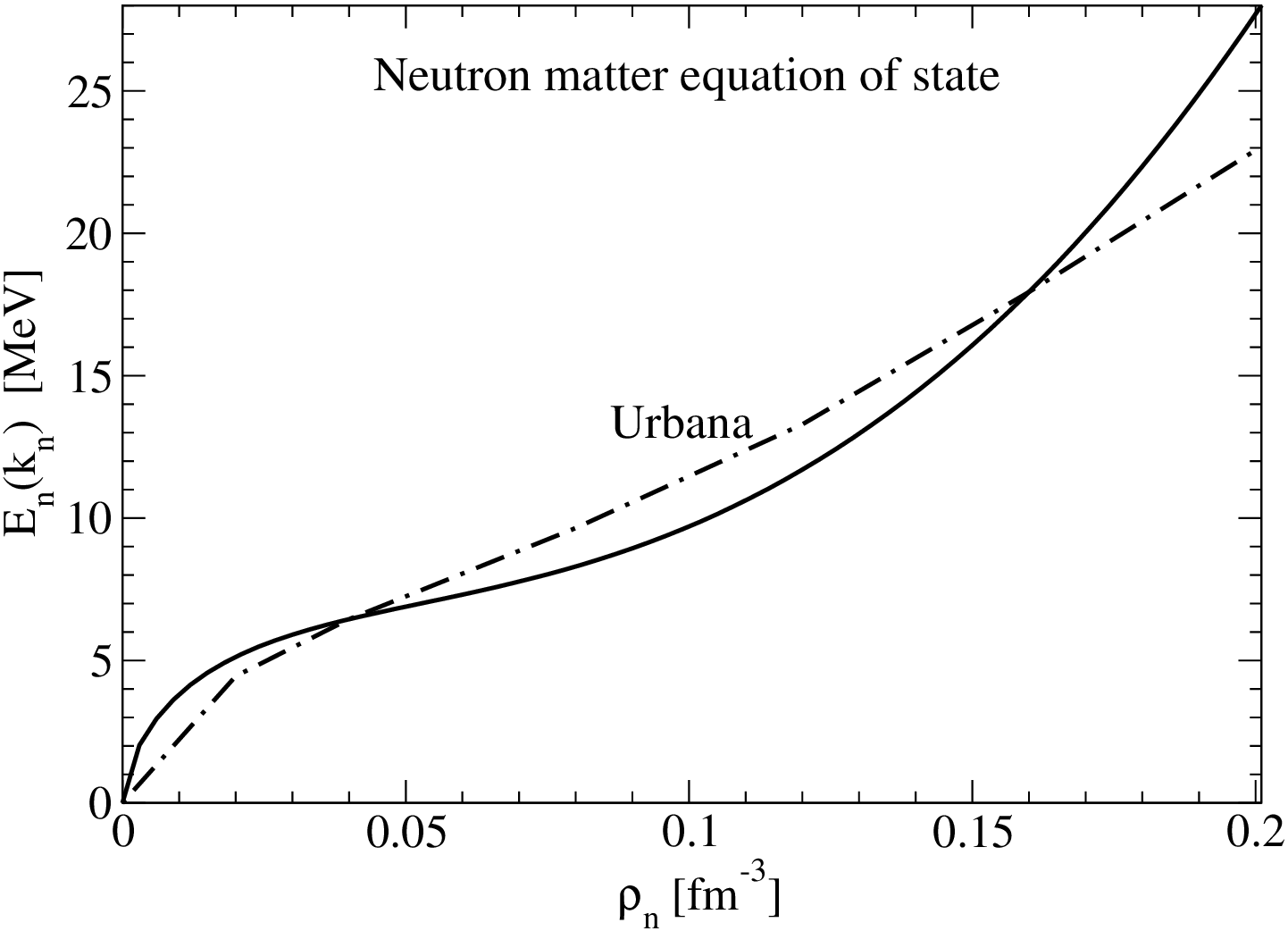}
\end{center}
\vspace{-0.4cm}
{\it Fig.\,4: Energy per particle $\bar E_n(k_n)$ of pure neutron matter as a 
function of the neutron density $\rho_n = k_n^3/3\pi^2$. It includes an
adjusted short-distance term of the form: $\bar E_n(k_n)^{(adj)} = -1.04\,$GeV$
^{-2}\,k_n^3 - 18.4 \,$GeV$^{-4}\,k_n^5$.  The dashed-dotted line stems from the 
sophisticated many-body calculation of the Urbana group \cite{akmal}.} 
\end{figure}

Finally, we turn to the behavior of the density-dependent quark condensate 
$\langle \bar q q \rangle(\rho_n)$ in the chiral limit $m_\pi \to 0$. As
demonstrated in Sec.\,II\,F of ref.\cite{homont} the (singular) chiral 
logarithms $\ln(m_\pi/\lambda)$ from irreducible $2\pi$-exchange and those
from the $m_\pi$-dependent vertex corrections to the $1\pi$-exchange cancel 
exactly in the case of isospin-symmetric nuclear matter. This subtle balance 
does not work anymore for pure neutron matter. The following singular piece 
remains: 
\begin{equation}
D_n(k_n)|_{m_\pi\to 0} =  {2k_n^3\over 3(4\pi f_\pi)^4}(1+6g_A^2-11 g_A^4) \ln
{m_\pi \over \lambda}\,. \end{equation} 
In order to understand the physical origin behind this infrared singularity 
let us consider the simplified situation with the $p$-wave pion-nucleon 
coupling set to zero: $g_A=0$. In that case only the $2\pi$-exchange
interaction generated by the Weinberg-Tomozawa term (at second order) 
survives. It gives rise to an isovector ($\sim \vec \tau_1 \cdot \vec \tau_2$) 
central NN-interaction with $m_\pi^2$-derivative equal to:  
\begin{equation} {\partial W_C(q) \over \partial m_\pi^2} = {1
\over 64 \pi^2 f_\pi^4} \Bigg\{ {1 \over 2} - \ln {m_\pi \over \lambda} -
{\sqrt{4m_\pi^2 + q^2} \over q} \ln { q+ \sqrt{4m_\pi^2 + q^2} \over 2m_\pi}
\Bigg\} \,. \end{equation} 
The two-body Hartree term (left diagram in Fig.\,5) in infinite neutron 
matter is proportional to its value at zero momentum transfer $q=0$, thus 
exhibiting the chiral logarithm, $-1/2 - \ln(m_\pi/\lambda)$. On the other
hand the limit $m_\pi\to 0$ of Eq.(10) exists and it is proportional to $1/2 - 
\ln(q/\lambda)$. Consequently, the two-body Fock term (right diagram in
Fig.\,5), involving the integral: $\int_0^{2k_n} dq\, q^2(2k_n-q)^2(4k_n+q)[1/2-
\ln(q/\lambda)]= 4k_n^6[5-4\ln(2k_n/\lambda)]/3$, stays finite in the chiral 
limit. The coordinate-space ``potential'' associated with the logarithm  
$\ln (q/\lambda)$ in momentum space actually behaves as $r^{-3}$ (for $r>0$). 
Its long-range character causes the interaction energy density for a 
homogeneous neutron-sphere of radius $R$ to grow logarithmically with the 
system size:
\begin{figure}
\begin{center}
\includegraphics[scale=0.9,clip]{wt2fig.epsi}
\end{center}
\vspace{-0.4cm}
{\it Fig.\,5: Three-loop Hartree and Fock diagrams generated by the
Weinberg-Tomozawa contact vertex at second order. Their combinatoric factor is 
$1/4$.}
\end{figure}
\begin{eqnarray} && {3\over 4\pi R^3}\!\int\limits_{|\vec r_{1,2}|<R}  \!\!\!\!
d^3r_1d^3r_2 \int {d^3 q \over (2\pi)^3} \, e^{i\vec q \cdot(\vec r_1-\vec  r_2)} \, \ln{
\lambda  \over q} \nonumber \\ && = {6\over \pi} \int_0^\infty dx \,{(x\cos x 
-\sin x)^2 \over x^4} \ln{\lambda R \over x} = \ln (2\lambda R) + \gamma_E
-{7\over 3} \,. \end{eqnarray} 
In other words, the usual the thermodynamic limit $N\to \infty,\, R\to\infty$, 
with $N/R^3 =$\,constant, does not exist for such a long-ranged interaction.
In this context the quantity $\langle \tau_3\rangle /f_\pi^2$ plays the role
of an isovector ``polarizability'' for the coupling and exchange of a pair of
massless pions which causes the infrared singular behavior. Since the
expectation value of $\tau_3$ does not vanish whenever the nuclear medium is 
asymmetric in isospin, the corresponding in-medium quark condensate will
encounter an infrared singularity in the chiral limit $m_\pi\to 0$.

In passing we note that the result for the quark condensate in pure proton 
matter (ignoring the Coulomb interaction) is the same as that for pure 
neutron matter, shown in Figs.\,1,\,3. Our results for the in-medium chiral 
condensate can also be interpreted in terms of a density-dependent effective 
nucleon sigma-term, $\sigma_{N,\rm  eff}(\rho)= \sigma^{(0)}_{N,\rm eff}(\rho)+\langle
\tau_3\rangle^2\,\sigma^{(1)}_{N,\rm eff}(\rho)$, with $\langle\tau_3 \rangle=(Z-N)/
(Z+N)$ the relative isospin-asymmetry. Its isoscalar piece $\sigma^{(0)}_{N\rm  eff}
(\rho)$ drops from the value $\sigma_N \simeq 45\,$MeV in vacuum to about 
$20\,$MeV at $\rho =0.35\,$fm$^{-3}$ (see Fig.\,7 in ref.\cite{homont}). The 
isovector piece $\sigma^{(1)}_{N,\rm eff}(\rho)$, which is maximally active in pure 
neutron or proton matter, brings it then back to almost the vacuum value in 
the density region $\rho = 0.35\,$fm$^{-3}$ considered here. As a consequence, 
the nuclear scalar mean-field associated in certain models 
\cite{finelli,qcdsum} with the in-medium condensate $\langle \bar q q\rangle(
\rho)$, would have an attractive isoscalar component as well as an attractive 
isovector component with approximately opposite density dependence. 

In summary, we have used in-medium chiral perturbation theory to calculate the
density-dependent quark condensate $\langle \bar q q \rangle(\rho_n)$ in pure 
neutron matter. We have found that the same $2\pi$-exchange dynamics which 
stabilizes the chiral condensate in isospin-symmetric nuclear matter
\cite{homont} does not alter (in a significant way) its linear decrease with 
the neutron density $\rho_n$. This different behavior originates from the 
reduced weight factors of the $2\pi$-exchange mechanisms in pure neutron
matter. We can conclude that the tendencies for chiral symmetry
restoration are actually favored in systems with large neutron excess
(e.g. neutron stars). Possible observable consequences of this feature should
be further investigated.

\end{document}